\documentclass[twocolumn,preprintnumbers,superscriptaddress,endnote,nofootinbib,aps,prd,floatfix]{revtex4}

\usepackage{datetime}
\usepackage{color}

\usepackage{footmisc,multirow}
\usepackage{subfigure}
\usepackage{amsmath}

\usepackage{graphicx}

\newcommand{\be}{\begin{eqnarray*}}
\newcommand{\ee}{\end{eqnarray*}}

\newcommand{\bee}{\begin{eqnarray}}
\newcommand{\eee}{\end{eqnarray}}
\newcommand{\beeq}{\begin{equation}}
\newcommand{\eeeq}{\end{equation}}

\newcommand{\ifb}{{\text{fb}}^{-1}}

\preprint{IPPP/15/73} \preprint{DCPT/15/73}

\begin{document}

\title{Measuring the Higgs-bottom coupling in weak boson fusion}

\begin{abstract}
  We study Higgs production through weak boson fusion with subsequent
  decay to bottom quarks.  By combining jet substructure techniques
  and matrix element methods in different limits we motivate this channel as a probe of
  the bottom-Yukawa interactions in the boosted regime.  In particular
  we ameliorate the ``no-go'' results of cut-and-count analyses
  in this channel. After applying a data-driven reconstruction approach we find 
  that the Higgs-bottom coupling can be limited to $0.82 < y_b/y_b^{\text{SM}} <1.14$ with $600~\mathrm{fb}^{-1}$.
\end{abstract}

\author{Christoph Englert} \email{christoph.englert@glasgow.ac.uk}
\affiliation{SUPA, School of Physics and Astronomy, University of
  Glasgow,\\Glasgow, G12 8QQ, United Kingdom\\[0.1cm]}

\author{Olivier Mattelaer} \email{o.p.c.mattelaer@durham.ac.uk}
\affiliation{Institute for Particle Physics Phenomenology, Department
  of Physics,\\Durham University, DH1 3LE, United Kingdom\\[0.1cm]}
  
\author{Michael Spannowsky} \email{michael.spannowsky@durham.ac.uk}
\affiliation{Institute for Particle Physics Phenomenology, Department
  of Physics,\\Durham University, DH1 3LE, United Kingdom\\[0.1cm]}

\maketitle


\section{Introduction}
\label{sec:intro}
After the Higgs discovery~\cite{hatlas,hcms} and a growing consistency
of Higgs measurements by ATLAS and CMS with the Standard Model (SM) 
hypothesis \cite{orig}, diversifying and extending Higgs to all
available production and decay channels is of utmost importance. On
the one hand, this strategy will us allow to over-constrain fits to, e.g.,
the dimension six extension of the Higgs sector, or, on the other
hand, could facilitate a new physics discovery in non-standard and less ``traditional'' Higgs
search channels.

The coupling of the Higgs boson to bottom quarks is
outstandingly important in this regard, because we expect a Higgs decay to bottom
final states at around 60\%~\cite{Dittmaier:2011ti}. Yet, none of the
currently available analyses is directly sensitive to this
coupling. Even the smallest deviation of the Higgs coupling to bottom
quarks has far reaching consequences for the Higgs lifetime; a modification of which 
might, e.g., point to a possible relation of the TeV scale with a hidden sector.  Since 
a modified Higgs phenomenology can arise from multiple sources,
fingerprinting the bottom Yukawa interaction is mandatory to
experimentally verify mass generation of the third generation down
sector, especially because the standard ways of looking for Yukawa
interactions such as Higgs production or bottom-quark associated Higgs
production suffer either from dominant virtual top-quark contributions
or a small total rate in light of a huge background. 

In fact, there are only a few processes that contribute to a direct measurement
of the bottom-Yukawa coupling: associated Higgs production  \cite{Butterworth:2008iy, Soper:2010xk, Butterworth:2015bya} and
top-associated Higgs production \cite{Plehn:2009rk,Artoisenet:2013vfa,Moretti:2015vaa}, with subsequent decay $H\to b\bar b$,
both of which are challenging to probe at the LHC, even with large
statistics. 

It is the purpose of this work to add another sensitive
channel to this list: weak boson fusion (WBF)-like Higgs production
with decay to bottom quarks. This channel has been studied in Ref.~\cite{Mangano:2002wn}, which
quoted a very small signal vs. background ratio, effectively removing
this process from the list of interesting Higgs processes. This is
mainly due to large backgrounds and little handle (such as a missing
central jet veto~\cite{gaporig,gaporig2,Barger:1991ar}) to control
them. In this work we extend the analysis of \cite{Mangano:2002wn} by employing novel reconstruction and all-information approaches
through combining shower deconstruction~\cite{Soper:2011cr, Soper:2012pb}, an all-order matrix element method to analyse fat jets, with the fix-order matrix element method
techniques~\cite{Kondo:1988yd, Abazov:2004cs} for the hard process. We show that the large backgrounds can be significantly
reduced, while a major part of the signal can be retained. This allows
us to ameliorate the no-go expectation of ``traditional''
cut-and-count analyses for WBF Higgs production with $b$-quark final
states.

This work is organised as follows. In Sec.~\ref{sec:evgen}, we comment on our event 
generation and the used analysis tools. Specifically, we review the matrix element method and shower deconstruction in Secs.~\ref{sec:mem} and 
\ref{sec:sd} to make this work self-contained.
Sec.~\ref{sec:results} is devoted to our results. We perform a naive cut-and-count analysis and show that kinematic handles alone do not provide enough discriminating power to sufficiently isolate signal from background. We show that the latter can be achieved with a combination of matrix element method and shower deconstruction techniques, leading to an expected sensitivity to the SM WBF contribution with around $100~\ifb$ luminosity. Sec.~\ref{sec:conc} provides a summary and gives our conclusions.

\section{Event generation and analysis tools}
\label{sec:evgen}

\subsection{Event generation}
\label{sec:events}

We generated events at Leading Order in the four flavor scheme with {\tt MadGraph5\_aMC@NLO} \cite{Alwall:2014hca,Degrande:2011ua,deAquino:2011ub}/{\tt Pythia8.2}  \cite{Sjostrand:2014zea}
  using NNPDF2.3 \cite{Ball:2012cx} for the parton distribution functions. 
The generation was split into five independent samples: two for the signal and three for the background. The two signal samples are the Higgs production in association with two light jets via either weak boson fusion (WBF) or via gluon fusion (GF) with the Higgs decaying into a $b\bar b$ pair. The gluon fusion process was generated via the new extension of {\tt MadGraph5\_aMC@NLO} supporting loop induced processes \cite{Hirschi:2015iia} and
includes the full top and bottom mass effects. For the background we split the $b\bar bjj$ final state into  pure QCD production (referred to as  $b\bar bjj$) and electroweak production (referred to as $Zjj$).
 The last background sample is the four light-flavor jet sample ($jjjj$), for which we limit ourselves to the pure QCD contribution. To avoid the double counting between the $jjjj$ and the $b\bar bjj$ samples related to $b$ emission in the parton-shower, we ran a four flavor parton-shower for the $jjjj$ sample.

At parton level a couple of loose cuts are applied in order to gain in efficiency. For all the samples with two $b$ quarks and two light jets in the final state, we apply the following cuts:
\begin{alignat*}{2}
p_{T,b} &\geq 20~\text{GeV} \,,\\
p_{T,j} &\geq 35~\text{GeV} \,,\\
y_{j_1} \cdot y_{j_2} & < 0 \,,\\
|y_{j_1} - y_{j_2}| &> 3.0 \,,\\
m_{j_1,j_2}  &\geq 500~\text{GeV}\,\\
o_{T, (b+\bar b)} &\geq 150~\text{GeV}\,,\\
\Delta R_{all, all} & \geq 0.2 \,. 
\end{alignat*}
For the $jjjj$ sample, the same cuts are applied with  the index ``$j_1,j_2$" being identified as the two most forward jets and the index ``$b$" refers to the two central jets.

\subsection{The Matrix Element method}
\label{sec:mem}

The matrix element method \cite{Kondo:1988yd, Abazov:2004cs} is based on the Neyman-Person Lemma \cite{NP_Lemma} stating that the best discriminant variable is the 
likelihood ratio where the likelihood is the product of probabilities evaluated on the sample. The probability of an event is computed in the matrix element method by calculating
\begin{equation*}
\mathcal{P_\alpha}(p^{exp}) = \frac{1}{\sigma}\int d\Phi(p^{part}) |M_\alpha(p^{part})|^2 P(p^{exp}|p^{part}),
\end{equation*}
where $p^{exp}$ represents the measured momenta for a given event, $p^{part}$ is the partonic phase-space point which we integrate over with a phase-space measure $d\Phi(p^{part})$ that also includes the parton distribution functions. $|M_\alpha(p^{part})|^2$ is the matrix element square for a given hypothesis $\alpha$ and $P(p^{exp}|p^{part})$, named the transfer functions, is the conditional probability to observe the experimental event under consideration for a given partonic phase-space point.

Using the best discriminant variable allows us to perform measurements for processes with extremely small cross-section or small event rate. However, even if the above integral can be computed via dedicated tools \cite{Artoisenet:2010cn}, this is very CPU intensive when performed for the full sample of events. In order to analyse large background samples efficiently, we therefore simplify the method by approximating the transfer function by a delta function, allowing us to drop the computation of the integral entirely \cite{Soper:2011cr, Andersen:2012kn}. This is conservative, since including such effect can only improve the sensitivity of the method\footnote{Further improvements could be achieved by evaluating the matrix elements at NLO accuracy \cite{Campbell:2012ct, Martini:2015zkx}.}. 

Therefore, for each event, the matrix element method is equivalent to computing the following likelihood ratio:
\begin{equation}
\label{eq:mem}
\chi_{MEM}= \frac{|M_{wbf}|^2 + |M_{gf}|^2}{|M_{jjjj}|^2+|M_{bbjj}|^2+|M_{Zjj}|^2}.
\end{equation}
For additional speed efficiency, the gluon fusion matrix element is not computed using the one loop matrix element -- like we did for the event generation -- but at tree level with an effective vertex coming for the integrating out the top quark loop \cite{Alwall:2007st}. The signal matrix elements ($GF$ and $WBF$) are computed for a three body final state (with the Higgs momentum being identified with the reconstructed $\bar{b}b$-pair momentum) while the backgrounds are computed for the four particle final state using the tagged $b$ subjet from the fat jet (see Sec.~\ref{sec:results}). To avoid potential bias, we use different sets of PDFs for the analysis (CT10 \cite{Guzzi:2011sv}) compared to the one used for the event generation.

\subsection{Shower Deconstruction}
\label{sec:sd}
Shower deconstruction \cite{Soper:2011cr, Soper:2012pb} is an all-order matrix element method designed to discriminate hadronically decaying electroweak-scale resonances, i.e. tops, $W/Z$ or Higgs boson, from QCD jets. 
First the constituents of a fatjet are reclustered into small inclusive jets, e.g. using the $k_T$ jet algorithm \cite{Ellis:1993tq} with R=0.2 and $p_{T,j} > 5$ GeV. One obtains a configuration of $N$ subjets with four-momenta $\{p\}_N = \{p_1, p_2, \cdots, \}$. Using these subjets as input to the method, a likelihood ratio is calculated from first-principle QCD, quantifying whether the observed distribution of subjets was initiated by the decay of a signal process, e.g. a Higgs boson, or background, e.g. a gluon. 
To calculate the likelihood ratio
\begin{equation}
\label{eq:sd}
\chi_{SD}(\{p\}_N) = \frac{P(\{p\}_N|S)}{P(\{p\}_N|B)},
\end{equation}
where $P(\{p\}_N|S)$ represents the probability of obtaining the subjet distribution $\{p\}_N$ given the signal hypothesis, and $P(\{p\}_N|B)$ is the probability for obtaining the same $\{p\}_N$ from background processes. To calculate $P(\{p\}_N|B)$ and $P(\{p\}_N|S)$ the method sums over all possible shower histories. In \cite{ATLAS:2014twa} it was shown that $\chi_{SD}$ is insensitive to pileup and shows good agreement between data and Monte-Carlo prediction.
We follow loosely the approach described in \cite{Soper:2014rya} to combine shower deconstruction with the fix-order matrix element method of Sec.~\ref{sec:mem}.

\section{Results}
\label{sec:results}
Based on the event generation detailed above, we first establish a baseline cut scenario inspired by \cite{Mangano:2002wn,Khoze:2003af}.

\begin{figure*}[t]
  \centering
 \includegraphics[width=8.0cm]{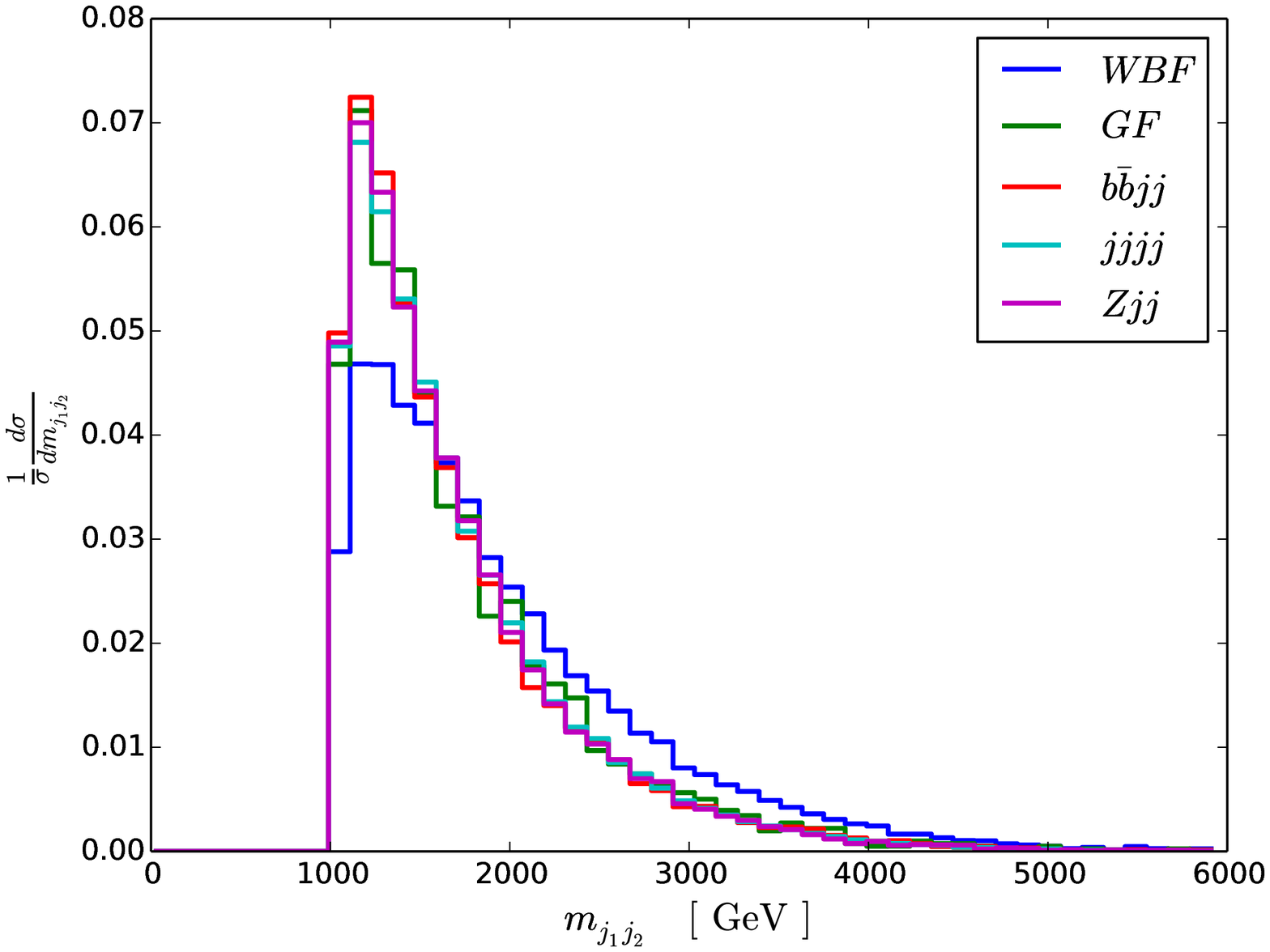}\hspace{1cm}
  \includegraphics[width=8.0cm]{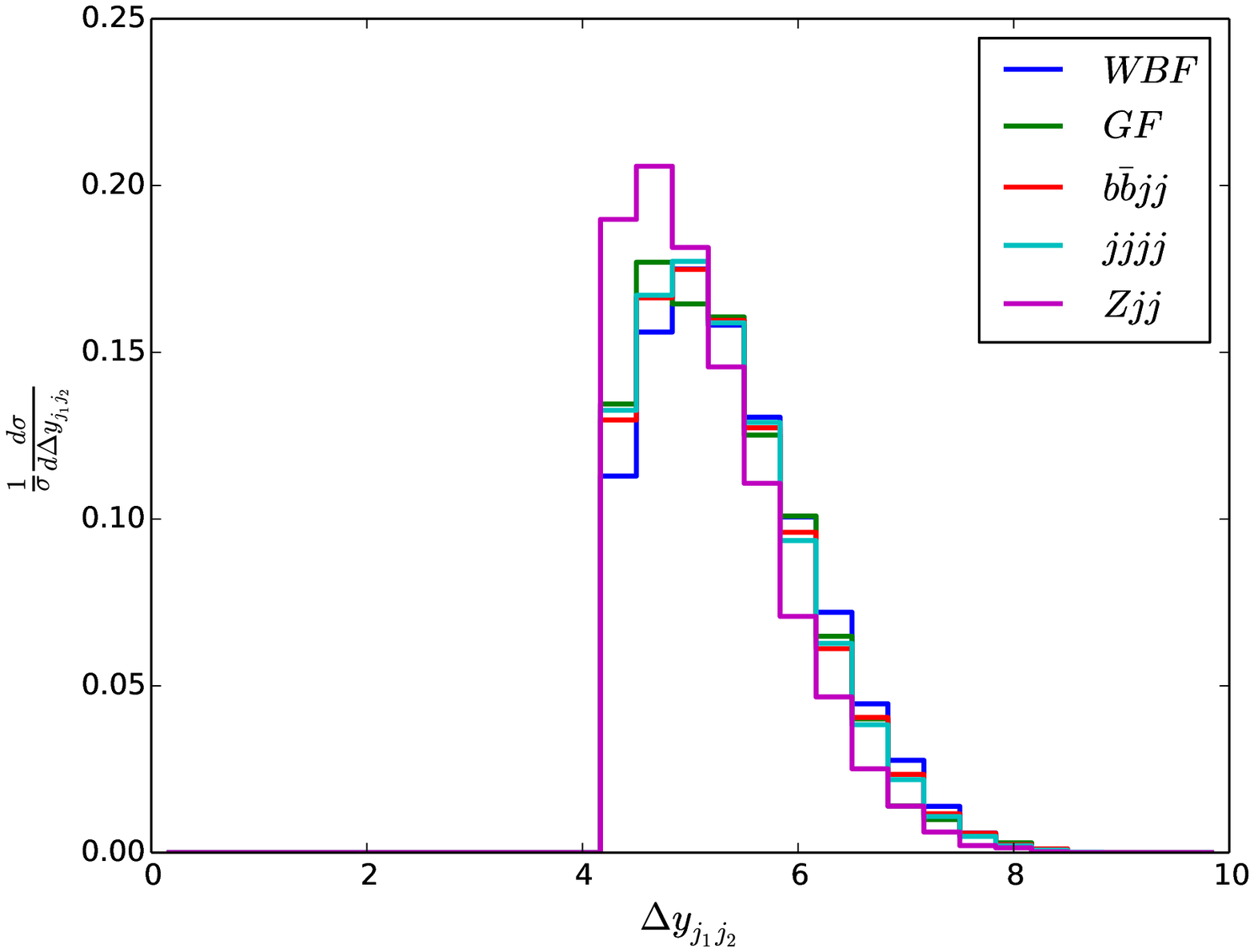}\\
 \includegraphics[width=8.0cm]{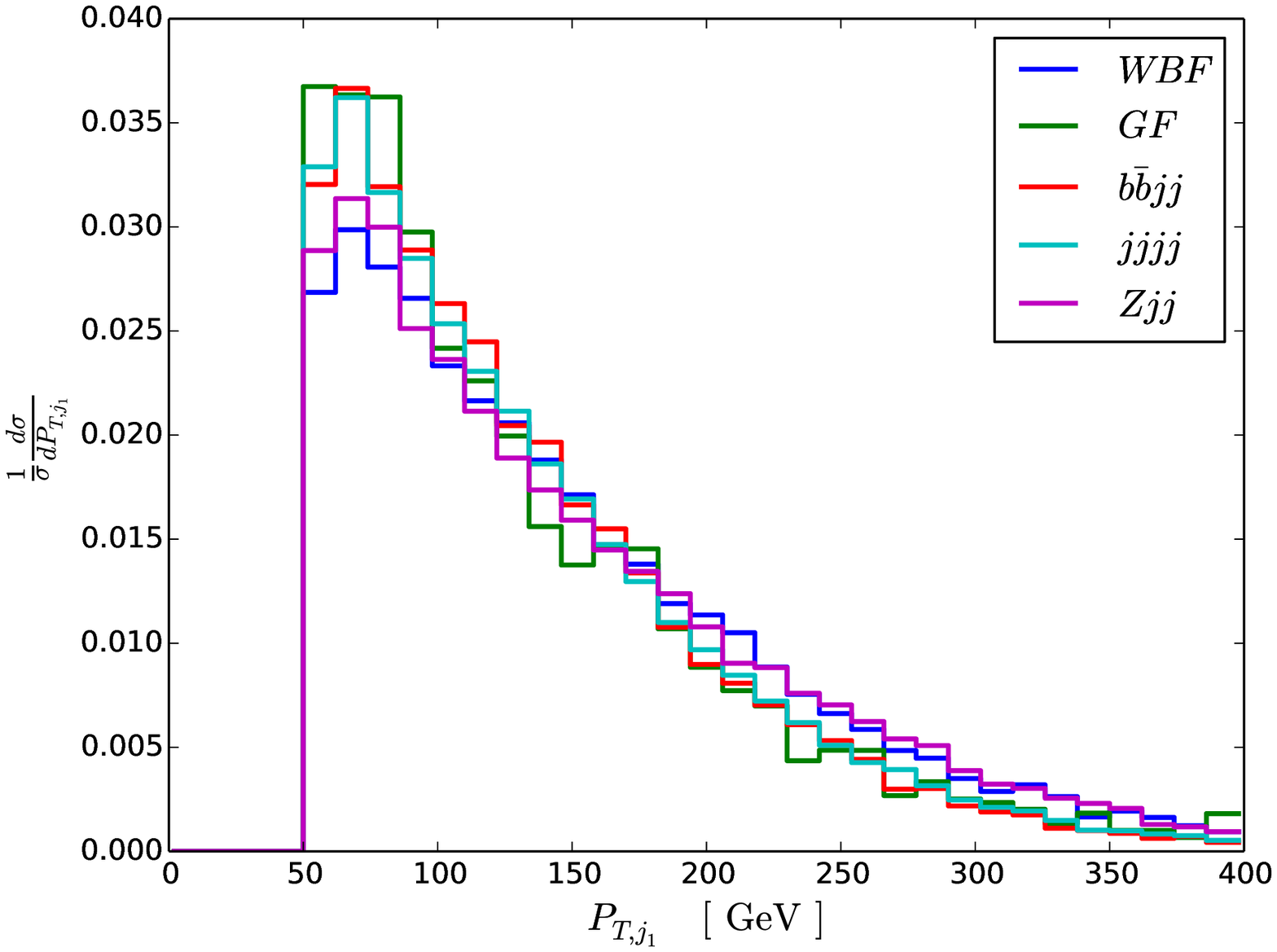}\hspace{1cm}
  \includegraphics[width=8.0cm]{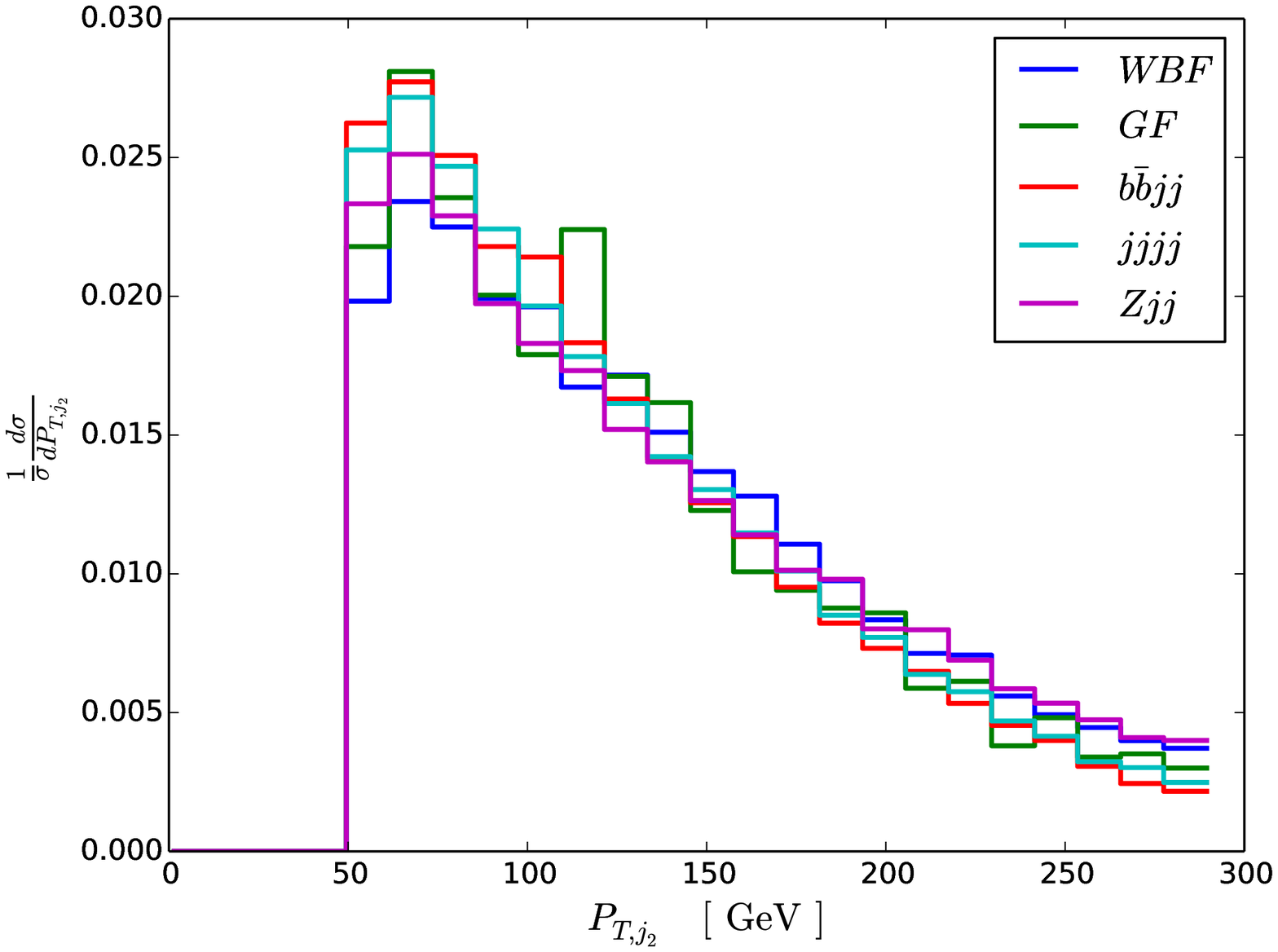}\\ 
  \caption{Representative kinematic distributions of signal and background processes after the analysis steps described in Sec.~\ref{sec:results} have been carried out.}
  \label{fig:basic}
\end{figure*}
\begin{table*}[!t]
  \begin{tabular}{l@{\quad}c@{\quad}c@{\quad}c@{\quad}|@{\quad}c@{\quad}|@{\quad}c@{\quad}|@{\quad}c@{\quad}}
          \hline
         &  & WBF & GF & $b\bar bjj$ & $Zjj$ & $jjjj$ \\
  \hline
  (i) & fat jet & 48.50 & 17.32 & 205109 & 553.16 & $2.23 \cdot 10^7$ \\[1ex]
  (ii) & wbf cuts  & 21.23 & 4.11 & 48441.9 & 127.98 & $5.18 \cdot 10^6$ \\[1ex]
  (iii) & mercedes star & 18.44 & 2.82 & 31674.5 & 84.975 & $3.39 \cdot 10^6$ \\[1ex]
  (iv) & fatjet $b$-tags & 4.59 & 0.578 & 3800.99 & 12.57 & 323.74\\[1ex]
  \hline
  \end{tabular}
  \caption{Cut flow of the cut-based analysis described in Sec.~\ref{sec:results}. The steps (i)-(iii) show the cross sections after the cuts of Eqs.~(\ref{eq:fatjet})-(\ref{eq:mstar}). Cross sections are quoted in units of femtobarns.}
  \label{tab:XS}
\end{table*}

In the first step, we veto events with isolated leptons with $|y_l| \leq 2.5$ and $p_{T,l}>10$ GeV. We then request a $R=1.2$
Cambridge-Aachen \cite{ca_algo} fat jet\footnote{Jet finding and clustering is
  performed with {\sc{FastJet}} \cite{Cacciari:2011ma}.} with
\begin{alignat}{2}
\label{eq:fatjet}
p_{T,{j_\text{fat}}} &> 200~\text{GeV} \, , \nonumber\\
|y_{j,\text{fat}}| & < 2.5 \,, \\
\mathrm{and}~~~ m_{j,\text{fat}} &> 90~\text{GeV} \, . \nonumber
\end{alignat}
After having identified a fat jet, we remove its
constituents from the final state, and the remaining constituents in the
event are clustered using anti-kT $R=0.6$ jets~\cite{antikt} with
$p_{T,j}>50$ GeV in $|y_{j}|<4.5$. The two jets with largest rapidity we define as
so-called tagging jets $j_1$ and $j_2$.

In the next step, we impose typical and stringent WBF selection requirements on these two tagging jets; they need to have a large
invariant mass, are required to lie in different detector hemispheres, and need to be separated by a large rapidity gap
\begin{alignat}{2}
\label{eq:wbf}
m_{j1,j2} &\geq 1000~\text{GeV} \,,\nonumber\\
y_{j_1} \cdot y_{j_2} & < 0 \,,\\
 |y_{j_1} - y_{j_2}| &> 4.0 \nonumber \,.
\end{alignat}
The first cut in particular decreases the gluon fusion contribution significantly which is a necessary requirement to end up with the clean
WBF-like selection to separate the impact of new physics between the two production modes \cite{Andersen:2012kn}. 

The typical WBF $pp \to h jj$ event topology that we want to isolate
is a Mercedes star configuration. Hence, we veto events for which one
of the tagging jets is central or collimated to the fat jet, i.e. we
require
\begin{alignat}{2}
\label{eq:mstar}
|y_{j_{1,2}}| &\geq 2.5 \nonumber \\
\Delta \phi(j_\text{fat},j_{(1,2)}) &> 2.0 
\end{alignat}
in the third step of the analysis. 

\begin{figure*}[!t]
  \centering
 \includegraphics[width=8.0cm]{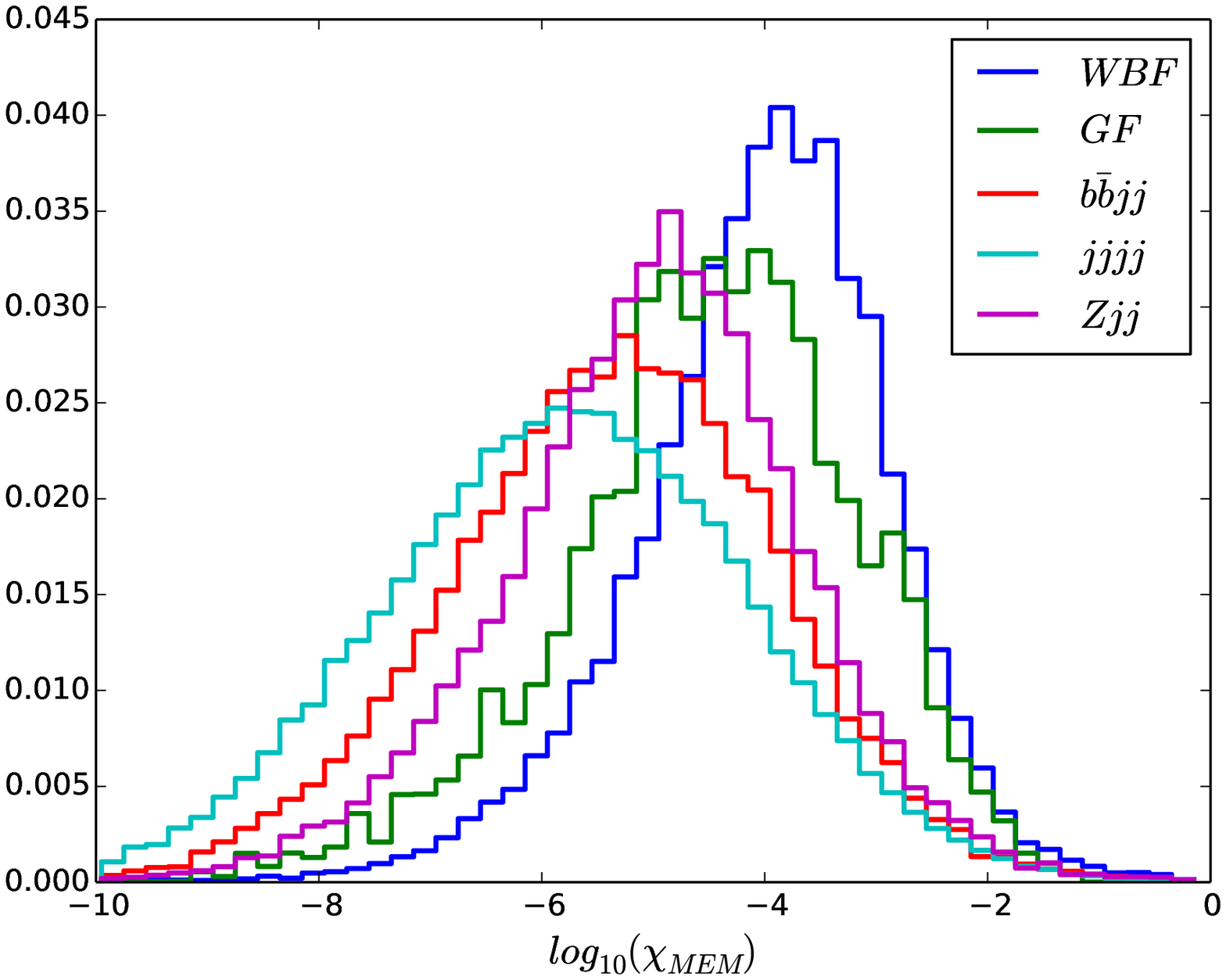}\hspace{1cm}
  \includegraphics[width=8.0cm]{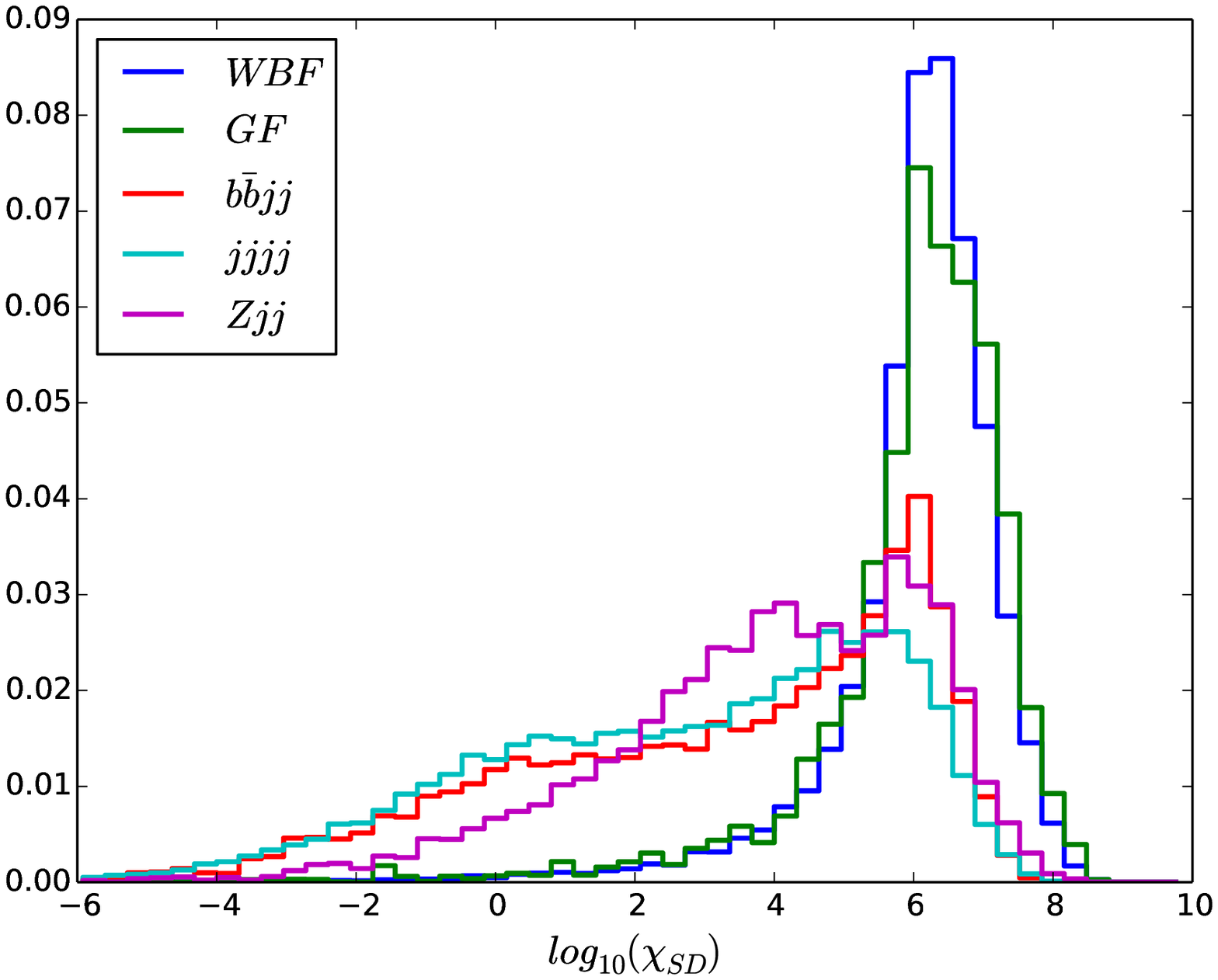}\\
  \caption{Distribution of events for both the matrix element method likelihood ratio (left) and shower deconstruction (right) after the analysis steps described in Sec.~\ref{sec:results} have been carried out.}
  \label{fig:likelihood}
\end{figure*}

Since the fat jet is produced centrally, we cannot use generic ways to
reduce the hadronic backgrounds, such as central/mini-jet vetos which
would also push the gluon fusion contribution to the percent level.

In the last step, we turn to the fat jet that we removed earlier from
the event record. We recombine the constituents of the fat jet to
so-called microjets with $R=0.2$ and $p_{T,j_m} \geq 5$ GeV. Of these
microjets we require the two with largest transverse momentum to be
$b$-tagged. We assume a flat $60\%$ tagging efficiency and $1\%$ fake
rate. To facilitate a $b$-tag, the microjets need to have $p_{T,j_m}
\geq15$ GeV. We decay the $B$-mesons, do not account for missing
energy in the reconstruction (hence our result is conservative), and,
in addition to the assumed tagging efficiency, for a $b$-tag we match
the $B$-meson to the respective jet by requiring $\Delta
R_{B_{\text{mes}},j} < R$.

A cut flow of our analysis is shown in Tab.~\ref{tab:XS};
distributions of signal and backgrounds are shown in
Fig.~\ref{fig:basic}. It is important to realise that after step (iii)
we have exhausted all kinematic handles to suppress the backgrounds
and find ourselves in the unfortunate situation that $b$-tagging is
not discriminative enough to cure a bad signal vs. background
ratio.\footnote{We have checked that the top pair background also
  contributes $\sim 3~\text{fb}$ at this stage, but becomes completely
  negligible in the matrix element method+shower deconstruction signal
  region detailed below.} Increasing the invariant mass cut would lead
to a further suppression of the backgrounds, however at an
unacceptably large decrease of the signal yield, leading to a
vanishing sensitivity to WBF.\footnote{Our results in this regard are
  compatible with the earlier results of Ref.~\cite{Mangano:2002wn}.}

\begin{figure}[!t]
 \includegraphics[width=8.0cm]{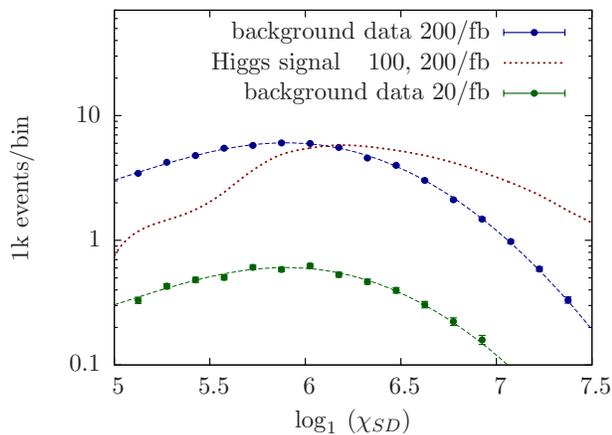}
 \caption{Expected background distribution of the shower
   deconstruction output for luminosities of $20~\ifb$ and
   $200~\ifb$. Note that the statistical uncertainty becomes
   negligible. For the $200~\ifb$ case we also compare the background
   fit to the Higgs signal distribution (multiplied by a factor
   100). Exploiting the different shape of the signal above a
   well-modeled background is key to the limit setting described in
   the text. The (asymmetric) error bars on each bin are
     calculated using quantiles following the ATLAS statistics
     recommendations \cite{cowan} inputting the expected background
     count.}
  \label{fig:datadriven2}
\end{figure}

\begin{figure}[!t]
 \includegraphics[width=8.0cm]{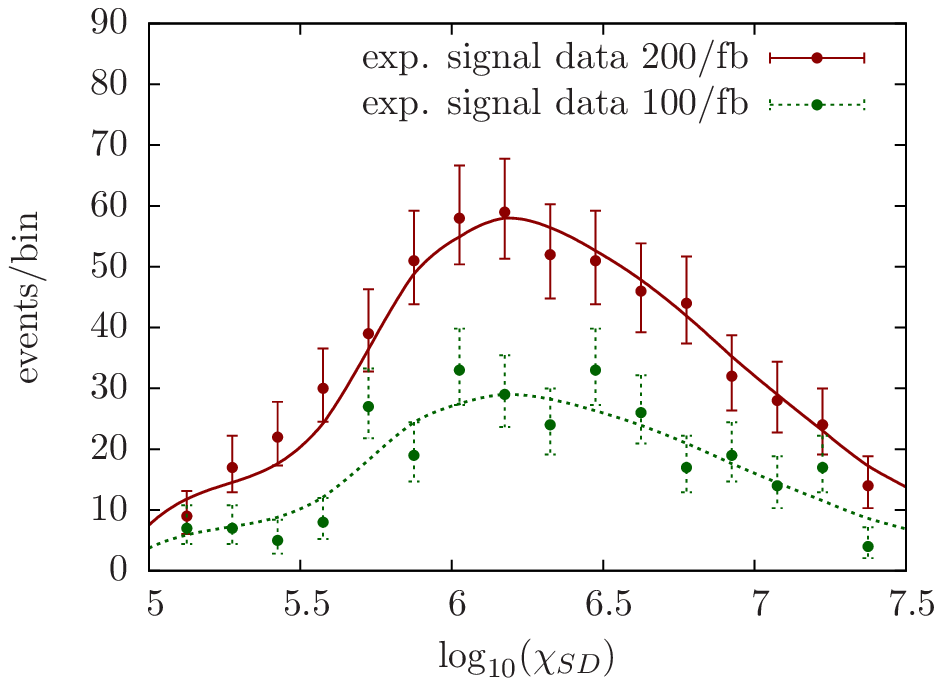}
 \caption{Expected signal distribution of the shower deconstruction
   output for $100$ and $200~\ifb$. We also plot signal
       pseudo-data to highlight the expected scatter of the signal
       events for the ideal scenario of a perfect background fit. The
       (asymmetric) error bars on each bin are calculated using
       the ATLAS statistics recommendations \cite{cowan} inputting the
       expected signal count.}
  \label{fig:datadriven3}
\end{figure}

With all ``traditional'' means exhausted we are left with two
approaches: (i) reverting to multivariate techniques such as boosted
decision trees or neural networks which heavily rely on the treatment
of systematics and training, combining a few observables that still
show promise to increase signal over background (see e.g. the recent
Ref.~\cite{Khachatryan:2015bnx}), or (ii) improve the analysis in a
way that is motivated by the asked physics question. The latter avenue
is exactly provided by combining the matrix element method for the
hard process with shower deconstruction, a matrix element method for
soft and collinear QCD radiation. The former provides a statistical
discrimination on the basis of expected signal and background
contributions, while the latter isolates the $H\to b\bar b$ decay from
the irreducible backgrounds by comparing the different shower
histories.

We separately calculate $\chi_{SD}$ of Eq.~\eqref{eq:sd} and
$\chi_{MEM}$ of Eq.~\eqref{eq:mem}, see
Fig.~\ref{fig:likelihood}. Again, this approach is conservative, as
the sensitivity of the analysis can be increased by combining shower
deconstruction and the matrix element method, see \cite{Soper:2014rya}
for a discussion. The two-dimensional correlation of matrix element
method and shower deconstruction isolates a region of phase-space with
a maximum signal-over-background ratio of $1/13.9$, requiring at least
3 signal events in isolated bins at 100/fb. Since shower
deconstruction can highly discriminate between $H\to b \bar b$ and the
continuum background, by integrating over adjacent bins in the shower
deconstruction observable, we obtain this way 10 events for 100/fb at
a reduced signal vs. background ratio of $1/28.8$. While this
constitutes a tremendous improvement over the cut and count analysis
with $S/B\simeq 1/800$, it is still obvious that background
systematics can have a significant impact.

Since this would render a likelihood analysis of the matrix element method-shower deconstruction correlation unreliable, we instead turn to a data driven approach based on pre-selecting a matrix element likelihood regime favored by the signal and fitting the background distribution of the shower deconstruction output. In practice, a fit based on a product of a Gaussian and a polynomial of degree 2 
\begin{equation}
\label{eq:fitfu}
f(x)= (ax^2 + bx + c)\exp\left[d (x - x_0)^2\right]
\end{equation}
with fitted parameters $a,b,c,d,x_0$ works very well as the background
template with $x=\log(\chi_{SD})$, see
Fig.~\ref{fig:datadriven2}. This opens the possibility to perform a
search for the Higgs boson similarly to the search for the Higgs in
other channels under the conditions that the background is not very
well understood, leading to large systematics once theoretical
uncertainties are treated at face value: one uses a background fit
function that is motivated from MC analysis to fit the distribution,
which then has small systematic uncertainty and looks for enhancements
over the expected background model. With increased statistics the fit
quality becomes better and systematic uncertainties become negligible
quickly (again similar to the situation in $H\to \gamma\gamma$
searches). Concretely we find the signal to be clustered around
$\log(\chi_{SD})\simeq 6$ while the background distribution is a
smooth Gaussian-like distribution of Eq.~\eqref{eq:fitfu} in the
search region $\log(\chi_{SD})>5$.

\begin{figure}[!t]
 \includegraphics[width=8.0cm]{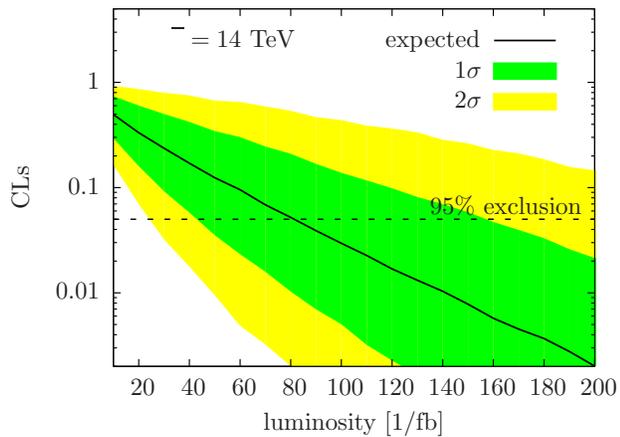}
 \caption{Expected exclusion using the CLs method \cite{cls} based on
   the data-driven analysis. For details see text.}
  \label{fig:datadriven}
\end{figure}

At a luminosity $\lesssim 100$ fb the total number of signal events,
as well as the different shape of the signal distribution,
Figs.~\ref{fig:datadriven2} and \ref{fig:datadriven3}, becomes
resolvable on the basis of the binned log likelihood method of
\cite{llhr, cls} (we stress again that the error bars in
Figs.~\ref{fig:datadriven2} and \ref{fig:datadriven3} are purely
exemplary; the correct combined background+signal distributions are
sampled in using the methods of \cite{llhr, cls} and taken into
account in the limits we quote). This means we can start excluding the
SM at 95\% confidence level in case the Higgs has a suppressed bottom
Yukawa interaction, Fig.~\ref{fig:datadriven}. On the other hand,
assuming a SM-like Higgs boson, we can turn this exclusion into a
coupling measurement. Projecting to $600~\ifb$, we obtain a constraint
\begin{equation*}
0.82 < y_b/y_b^{\text{SM}} <1.14
\end{equation*}
at 95\% confidence level by propagating the impact of the modified
bottom Yukawa interaction through the Higgs production and decay
phenomenology while keeping all other Higgs couplings fixed to their
SM values.

\section{Summary and Conclusions}
\label{sec:conc}
In this paper we have performed an analysis of Higgs production via
weak boson fusion (WBF) with subsequent decays $H\to b\bar b$.  While
this process is heavily plagued with backgrounds due to the
non-availability of signal vs. background enhancing strategies like
the central jet veto, we have shown that by combining novel analysis
strategies, we can elevate the discouraging result of a simple
cut-and-count analysis that exploits the basic kinematic features of
WBF to a sensitive strategy at a luminosity of about $100~\ifb$.  Our
strategy has the additional advantage that it relies on a data-driven
fit of the background template, which is derived from first principle
QCD and fixed-order calculations instead of relying on nontransparent
multivariate techniques. Crucial to this strategy is that the matrix
element method and shower deconstruction combine complementary
information - an analysis solely based on one of these tools does not
provide enough discriminating power to increase the sensitivity to WBF
Higgs production in the bottom final state. As a result, during the
upcoming run of the LHC, ATLAS and CMS will be able to probe the
Higgs-bottom coupling in a complementary way to well established
measurements in $\bar{t}tH$ and $VH$.


\acknowledgements M.S. acknowledges financial support by HiggsTools
ITN under grant agreement PITN-GA-2012-316704.
O.M. is a Durham International Junior Research
Fellow.


\end{document}